# Mimicking cuprates: large orbital polarization in a metallic square-planar nickelate

Junjie Zhang,[1,*] A. S. Botana,[1] J. W. Freeland,[2] D. Phelan,[1] Hong Zheng,[1] V. Pardo,[3,4] M. R. Norman,[1] J. F. Mitchell[1]

[1]Materials Science Division, Argonne National Laboratory, Argonne, Illinois 60439, USA

[2]Advanced Photon Source, Argonne National Laboratory, Argonne, Illinois 60439, USA

[3]Departamento de Fisica Aplicada, Universidade de Santiago de Compostela, E-15782 Santiago de Compostela, Spain

[4]Instituto de Investigacions Tecnoloxicas, Universidade de Santiago de Compostela, E-15782 Santiago de Compostela, Spain

**Abstract**

High temperature cuprate superconductivity remains a defining problem in condensed matter physics. Among myriad approaches to addressing this problem has been the study of alternative transition metal oxides with similar structures and $3d$ electron count that are suggested as proxies for cuprate physics. None of these analogs has been superconducting, and few are even metallic. Here, we report that the low-valent, quasi-two-dimensional trilayer compound, $Pr_4Ni_3O_8$ avoids a charge-stripe ordered phase previously reported for $La_4Ni_3O_8$, leading to a metallic ground state. By combining x-ray absorption spectroscopy and density functional theory calculations, we further find that metallic $Pr_4Ni_3O_8$ exhibits a low-spin configuration and significant orbital polarization of the unoccupied $e_g$ states with pronounced $d_{x^2-y^2}$ character near the Fermi energy, both hallmarks of the cuprate superconductors. Belonging to a regime of $3d$ electron count found for hole-doped cuprates, $Pr_4Ni_3O_8$ thus represents one of the closest analogies to cuprates yet reported and a singularly promising candidate for high-$T_c$ superconductivity if appropriately doped.

A collection of structural, magnetic and electronic characteristics are widely considered to be key ingredients in high temperature cuprate superconductivity: a quasi-2D square lattice, spin-½, strong antiferromagnetic correlations, large orbital polarization of the unoccupied $e_g$ states (with a single $d_{x^2-y^2}$ band near the Fermi energy), and a plethora of broken symmetry phases[1,2] (*e.g.*, charge stripes and spin density waves) proximate to the parent insulating phase. One design strategy for finding new high-$T_c$ materials has been to build solids that possess some or all of these electronic and structural features in hopes that superconductivity will emerge.[3-5] For example, the 5$d$ transition metal oxide $Sr_2IrO_4$, with a $J_{eff}$ = ½ ($t_{2g}^5$ valence shell) ground state, has recently been found to satisfy many of the above-mentioned criteria and predicted to show superconductivity.[6] Angle-resolved photoemission spectroscopy measurements have shown both Fermi arcs[5] and the opening of a *d*-wave gap[7] (also seen by scanning tunneling spectroscopy[8])—both hallmarks of cuprate superconductors—upon surface doping with electrons. However, no evidence for superconductivity exists, even for doped samples.[5,7,8] Due to its proximity to copper in the periodic table, nickel-based oxides have long been considered as potential cuprate analogs, and indeed tantalizing features such as Fermi arcs have been observed in the metallic but non-superconducting nickelate $Eu_{0.9}Sr_{1.1}NiO_4$.[9] The potential for nickelate superconductivity has been the subject of much theoretical discussion.[4,10,11] Anisimov *et al.* have argued that square-planar coordinated $Ni^{1+}$ (isoelectronic with $Cu^{2+}$, $d^9$) doped with $S=0$ $Ni^{2+}$ holes will host superconductivity,[10] and Chaloupka and Khaliullin have speculated that the low-energy electronic states of $LaNiO_3/LaMO_3$ (M=Al, Ga…) heterostructures can be mapped to the single-band model as developed for cuprates.[4] In contrast, Lee and Pickett have argued that weak *p-d* hybridization in $LaNiO_2$, resulting from an energy mismatch between Ni 3$d$ and O 2$p$ orbitals, weakens the analogy between nickelates and cuprates.[11] Here, by combining x-ray absorption spectroscopy (XAS) and density functional theory (DFT) calculations, we demonstrate that the trilayer nickelates $R_4Ni_3O_8$ (R=La, Pr, hereafter R438) express key electronic signatures of cuprates:

a large orbital polarization with unoccupied Ni 3$d$ states that are predominately $d_{x^2-y^2}$ in character, a low-spin configuration for dopants (Ni$^{2+}$ holes), and a high degree of hybridization between Ni 3$d$ and O 2$p$ states. We further show that Pr438 avoids the charge-ordered stripe phase found in La438,[12] leading to a metallic ground state possessing a highly orbitally polarized electronic configuration. These findings argue that the R438 family, and in particular Pr438, lies closer to superconducting cuprates than previously studied nickelates such as La$_{2-x}$Sr$_x$NiO$_4$ and LaNiO$_3$-based heterostructures.

The trilayer nickelates R438 are formally Ni$^{1+}$ compounds that are self-doped with Ni$^{2+}$ (nominally Ni$^{1+}$:Ni$^{2+}$=2:1) to yield an average Ni valence of 1.33+, *i.e.*, ⅓-hole doped into the Ni$^{1+}$ background. Their quasi-2D structure[12] can be described as three consecutive layers of corner-sharing NiO$_2$ square planes alternating with R$_2$O$_2$ fluorite-type layers, stacked along the $c$ axis (Fig. 1$a$). With 8.67 $d$ electrons per Ni on average, the $e_g$ states may be energetically arranged in two different ways (assuming the $t_{2g}$ states are fully occupied): the limit where the Hund's rule coupling is larger than the crystal field splitting between the two $e_g$ orbitals (high-spin state, HS) or the opposite situation (low-spin state, LS), as shown in Fig. 1$b$. While information on Pr438 is limited only to a remark that it is a black semiconductor,[13] La438 is well-known to undergo a semiconductor-insulator transition at 105 K.[12,14-17] Three different mechanisms have been proposed to explain this transition: a spin density wave nesting instability of the Fermi surface,[14] a change of the Ni spin state,[15,16] and more recently the formation of charge stripes.[12,17]

Single crystals of La438 and Pr438 were prepared by reducing their parent n=3 Ruddlesden-Popper (R-P) compounds, R$_4$Ni$_3$O$_{10}$ (R=La, Pr, hereafter R4310, see Fig. 1$a$ and Supplementary Fig. 1). To our knowledge, this is the first report of single crystal growth of Pr4310 and Pr438 (see Supplementary Table 1,2 for structural details of Pr438). The valence of Ni in R438 and R4310 is consistent with the expected values 1.33+ and 2.67+, respectively, as measured by Ni $L$ edge spectroscopy (Supplementary Fig. 2).

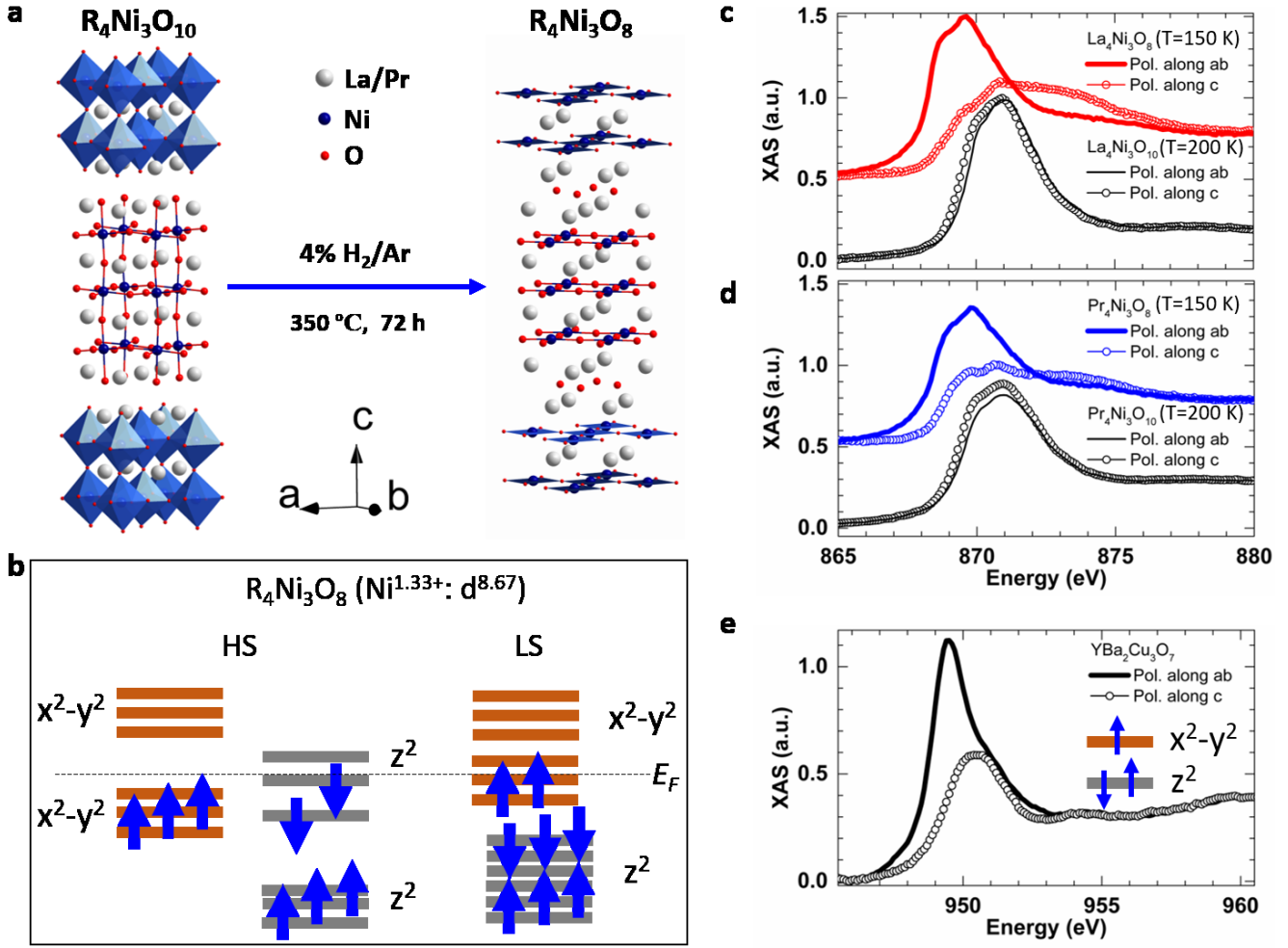

**Figure 1. Orbital polarization of $R_4Ni_3O_8$ (R=La, Pr).** (*a*) Crystal structures of the parent Ruddlesden-Popper phase $R_4Ni_3O_{10}$ (R=La, Pr) and of the square-planar phase $R_4Ni_3O_8$ (R=La, Pr) obtained by hydrogen reduction. (*b*) Energy level diagrams of $e_g$ states for proposed high-spin (HS) and low-spin (LS) Ni in $R_4Ni_3O_8$ (Ref. 15, 16). Three Ni atoms per formula unit are represented. $E_F$ denotes the Fermi energy. (*c, d*) Polarization dependence of the X-ray absorption spectroscopy (XAS) of $R_4Ni_3O_{10}$ at 200 K and $R_4Ni_3O_8$ at 150 K with data collected in the total fluorescence yield (TFY) mode. (*e*) Polarization dependence of XAS of $YBa_2Cu_3O_7$ in the TFY mode (Ref. 19). Inset: energy level diagram of $e_g$ states for cuprates (per copper).

Fig. 1c,d compares the polarization dependence of the XAS at the $L_2$ edge for R438 at 150 K and the corresponding parent compounds, R4310. The Ni $L_2$ edge was utilized due to the strong overlap between La $M_4$ and Ni $L_3$, which obscures the spectral features.[18] A large orbital polarization of the unoccupied $e_g$ states of R438 is immediately recognized by the strong intensity difference between in-plane (Pol. along

*ab*) and out-of-plane (Pol. along *c*) polarization near the leading part of the edge. This notable difference resembles that observed in high-$T_c$ cuprates, for example YBa$_2$Cu$_3$O$_7$,[19] as shown for comparison in Fig. 1e, which are considered to exhibit empty states of pure $d_{x^2-y^2}$ character. The XAS at the leading edge can hence be attributed to orbitals oriented in the *ab* plane (the unoccupied Ni $d_{x^2-y^2}$), which is consistent with the LS state picture shown in Fig. 1b. The data indicate that the out-of-plane oriented orbitals are largely occupied, although the intensity seen in the range $868 \leq E \leq 876$ eV implies some contribution from unoccupied $d_{z^2}$ states. We can quantify the orbital character utilizing a sum rule analysis,[18] which allows us to determine the ratio of the unoccupied orbital occupations. Specifically, the quantity $r = h_{z^2}/h_{x^2-y^2}$, where h denotes holes in a particular $e_g$ orbital, can be determined by an integration of the background-subtracted XAS data (see Supplemental Fig. 3 and Quantitative evaluation of the orbital polarization). Following this approach, we find that the R438 compounds show an $r < 0.5$, which indicates an orbital polarization favoring holes with $d_{x^2-y^2}$ character. This result agrees closely with the $r \sim 0.4$ derived from the theoretical density of states and represents ~70% of the holes in the $d_{x^2-y^2}$ orbital. For comparison, a similar analysis of YBa$_2$Cu$_3$O$_7$ yields an $r = 0.35$. We also note that the states contributing to the first 2 eV of the XAS yield an even smaller $r$ value, which indicates that the orbital polarization near the Fermi level is biased even more toward $d_{x^2-y^2}$ character.

We attribute this large orbital polarization in R438 to the square-planar environment of the Ni atoms, which leads to a large crystal-field splitting of the $e_g$ states, with $d_{x^2-y^2}$ orbitals expected to lie higher in energy than $d_{z^2}$. Comparing the orbital polarization of R438 with that of other nickelates that still retain the apical oxygens reveals more clearly the impact of the reduced coordination number of Ni. In sharp contrast to R438, the parent compounds R4310 give $r$ values slightly larger than one, which indicates a small orbital polarization in favor of $z^2$ holes. The single-layer R-P nickelates La$_{2-x}$Sr$_x$NiO$_4$ show negligible splitting at the absorption edge between in-plane and out-of-plane polarization, reflecting a small orbital polarization of the $e_g$ states.[20] By utilizing strain, charge transfer, and confinement, sizable orbital polarization at interfaces has been engineered in LaNiO$_3$-based heterostructures, which also contain NiO$_6$ octahedra.[21]

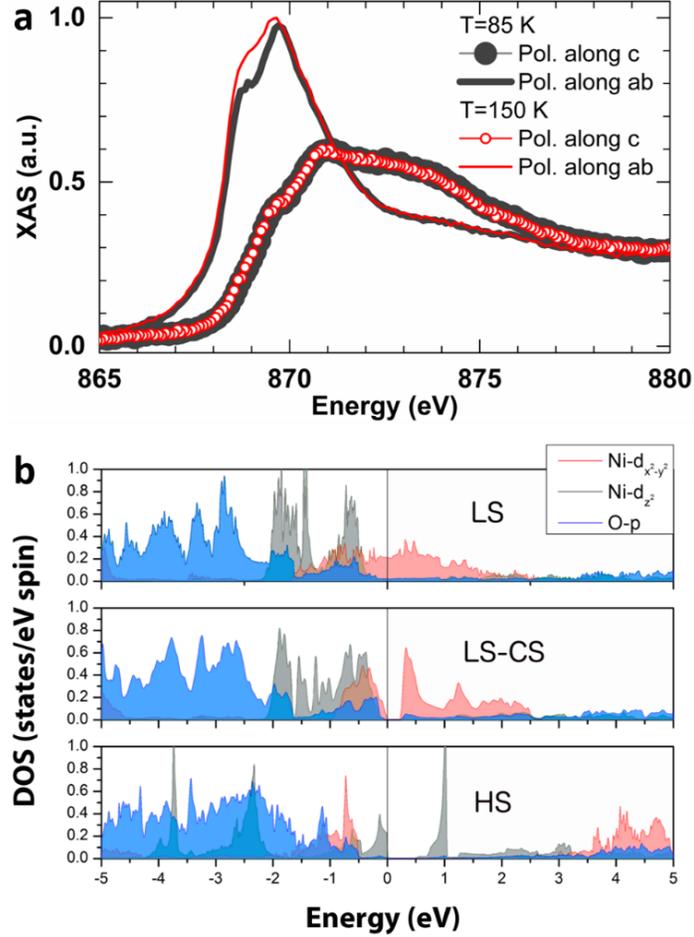

**Figure 2**. **Evidence for low-spin state nature of La$_4$Ni$_3$O$_8$.** (a) $c$ axis and $ab$ plane polarized Ni $L_2$ edge XAS of La$_4$Ni$_3$O$_8$ above and below the 105 K transition. (b) Ni-$e_g$ and O-$p$ orbital resolved density of states (DOS) for the low spin (LS), low-spin charge stripe (LS-CS) and high spin (HS) states from DFT calculations.

We now turn to the nature of the spin state below the 105 K transition in La438. As mentioned above, this transition has been attributed to a change in the Ni spin state,[15] and more recently it has been ascribed to the formation of charge stripes.[12] DFT calculations have found a Ni$^{1+}$/Ni$^{2+}$ charge stripe (CS) ground state with the Ni$^{2+}$ ions in a LS state ($S=0$) yielding an antiferromagnetic arrangement of Ni$^{1+}$ ($S=½$) ions (from here on denoted as the LS-CS state).[17] Fig. 2a shows the polarization dependent XAS of the Ni $L_2$ edge above and below the transition. Notably, across the transition, there is no change in the out-of-plane polarized XAS and only a slight change in the in-plane XAS. The $L_2$ leading edge is associated with the unoccupied Ni $d_{x^2-y^2}$ orbitals, consistent with a LS to LS-CS scenario but not with a LS-HS transition. In both LS and LS-CS states, as can be seen in Fig. 2b, the unoccupied Ni-3d states are calculated to be solely $d_{x^2-y^2}$ in character. Furthermore, in agreement with the XAS data, DFT calculations show no change in the total $d_{z^2}$ occupation and only a minor change in the $d_{x^2-y^2}$ occupation across the transition due to the opening of a small gap (0.25 eV in a calculation without a Hubbard $U$) in the LS-CS phase (Supplementary Fig. 4).

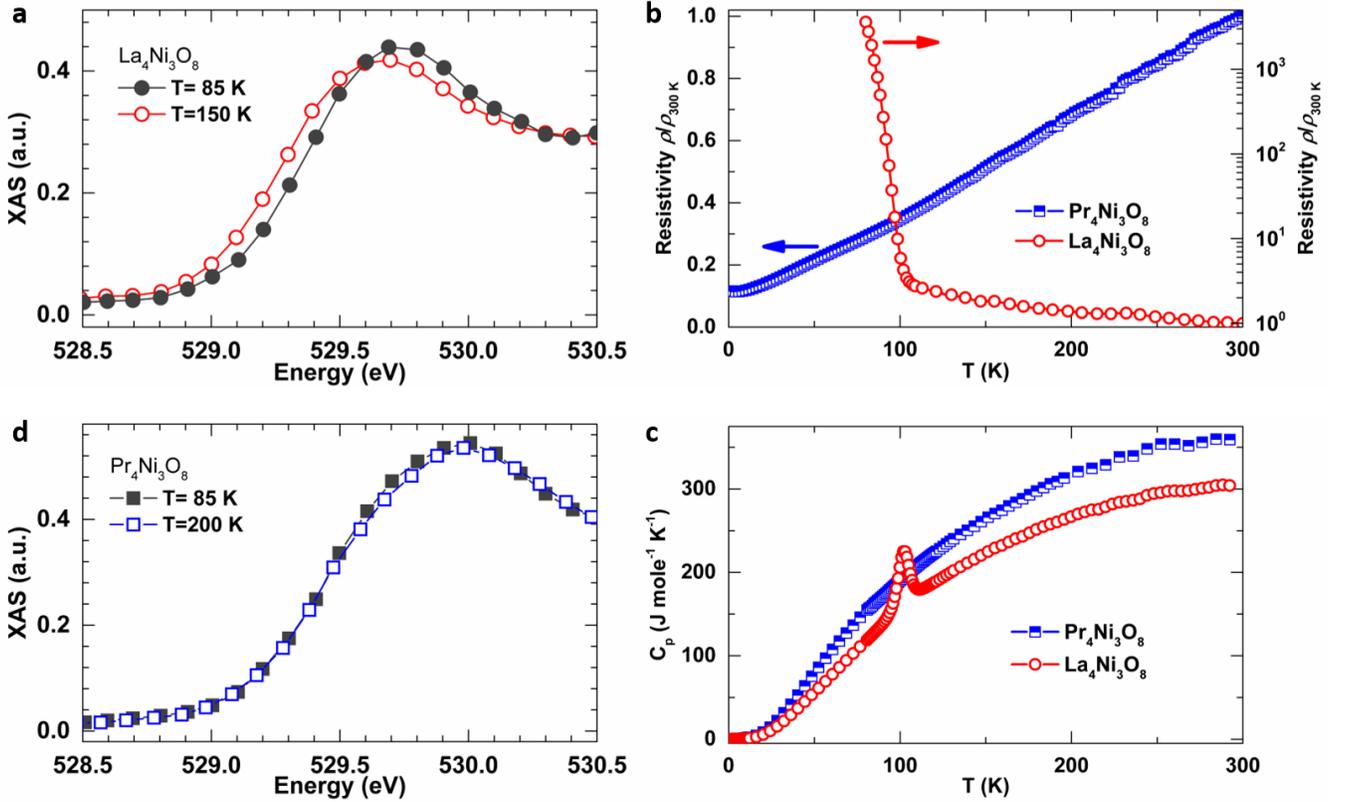

**Figure 3**. **Oxygen $K$ edge XAS and physical properties of La$_4$Ni$_3$O$_8$ and Pr$_4$Ni$_3$O$_8$.** (*a*) The pre-peak in the oxygen $K$ edge XAS for La$_4$Ni$_3$O$_8$ above and below 105 K. (*b*) Resistivity in the *ab* plane. (*c*) Heat capacity. (*d*) The pre-peak in the oxygen $K$ edge XAS for Pr438 at 85 K and 200 K.

We also measured the hybridization between Ni $d_{x^2-y^2}$ and O $2p$ states in the vicinity of the Fermi level. Fig. 3*a* shows the oxygen *K* edge XAS for La438 above and below the phase transition in the range of 528.5-530.5 eV. Like the hole-doped single layer nickelates and high-$T_c$ cuprates,[22] a well-defined pre-peak in La438 is due to holes located in the O $2p$ orbitals hybridized with the Ni $d_{x^2-y^2}$ states. This result agrees with DFT calculations demonstrating that in both LS and LS-CS states there is significant hybridization between O $2p$ and planar Ni $d_{x^2-y^2}$ states in the vicinity of the Fermi level (see Fig. 2*b*). This is a consequence of the strong interaction between Ni $d_{x^2-y^2}$ and O $2p$ states, since the lobes of the *d*-orbitals are directed along the planar Ni-O bonds. A small but clear shift to higher energy of the pre-peak reflects the opening of a small (~ 0.1 eV) gap in the CS phase, also consistent with our DFT calculations.

Thus far, we have shown that La438 shares several key features with superconducting cuprates: a large orbital polarization of the unoccupied $e_g$ orbitals, Ni$^{2+}$ ions in a LS state (in sharp contrast to HS in hole-doped single layer nickelates), and a $d_{x^2-y^2}$ orbital character near the Fermi energy with strong hybridization of Ni $3d$ and O $2p$ states. All of these features can be traced back to the square planar geometry of Ni with its $D_{4h}$ crystal field and missing apical oxygen ligands. Nevertheless, as a ⅓-hole self-doped material, La438 forms charge stripes below the 105-K transition to become an insulator, much like the analogous single layer nickelates, whereas the ⅓-hole doped cuprates examined so far are metals.[1,23] We now demonstrate that it is possible to establish a metallic phase in the R438 system while maintaining predominantly $d_{x^2-y^2}$ orbital polarization at a carrier concentration relevant to hole doped cuprates.

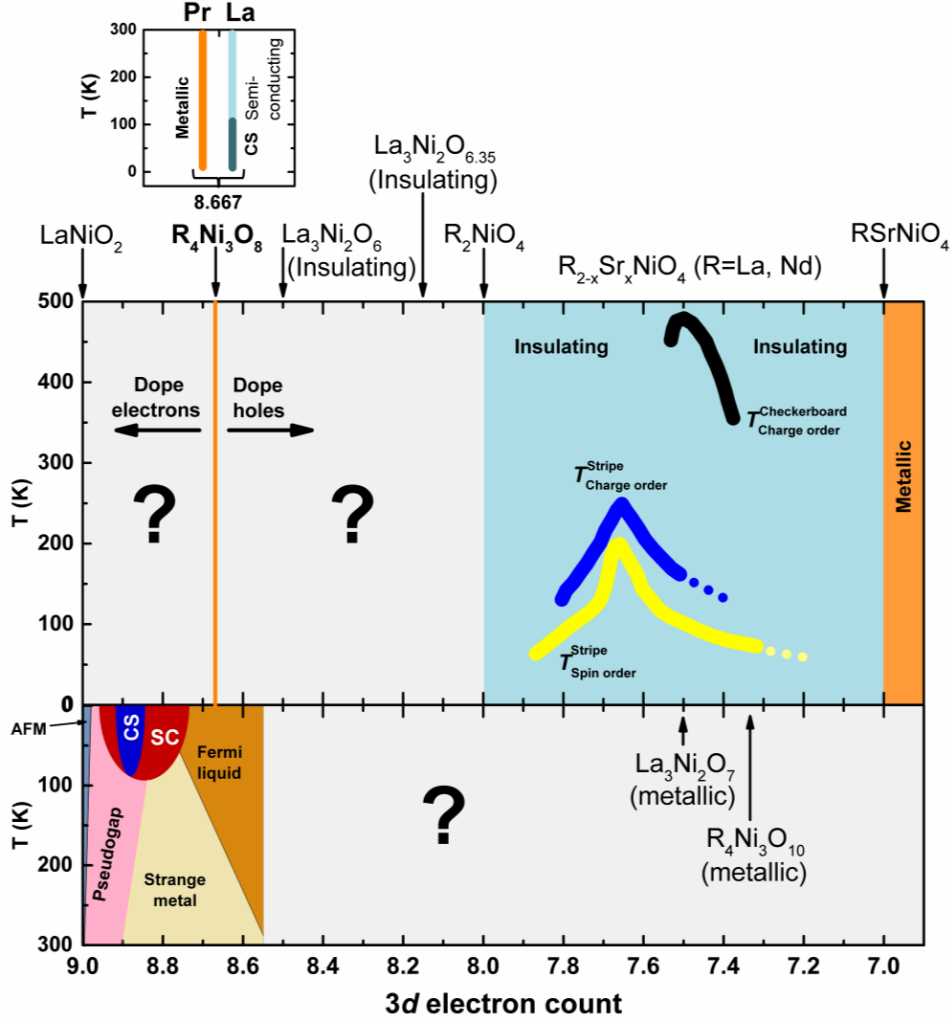

**Figure 4. Schematic phase diagram of layered nickelates and cuprates presented as a function of the nominal 3d electron count**. SC: superconductor; CS: charge stripe order; AFM: antiferromagnetic insulator. (Bottom) Generic phase diagram of the hole doped cuprates (Ref. 1, 23, 29). (Top) Electronic phase diagram of layered nickelates. Only a few line compositions are known between $d^9$ and $d^{6.9}$. Variable doping in this range has been achieved for single-layer Ruddlesden-Popper phases $R_{2-x}Sr_xNiO_4$ (R=La and Nd, Ref. 25, 26), with an insulator-metal transition at ~$d^7$. An expanded region around $d^{8.667}$ shows the electronic behavior of $R_4Ni_3O_8$ (R=La, Pr).

Fig. 3b presents the resistivity of a Pr438 single crystal, exhibiting a metallic temperature dependence, $d\rho/dT > 0$,[24] in the temperature range of 2-300 K, a distinct contrast with both La438 and the hole-doped single-layer R-P nickelates,[25] the latter of which only become metallic above a very large hole concentration $x \sim 1$ (see Fig. 4).[26] In addition, heat capacity (Fig. 3c) and crystal structure measurements (Supplemental Fig. 5) evidence no signature of a phase transition in Pr438 in the temperature range studied. Fig. 3d shows the oxygen $K$ edge XAS data for Pr438 at 85 and 150 K. Compared with La438, no shift in the pre-peak is found in Pr438, confirming that no gap opens from charge stripe formation in this temperature window. Finally, high-resolution single crystal x-ray diffraction data show no superlattice peaks (Supplemental Fig. 6) for temperatures as low as 15 K. Taken together, these findings argue that Pr438 does not form the static charge stripe phase found in La438, but remains metallic down to 2 K.

One plausible explanation for metallicity in Pr438 is a chemical pressure effect — the ionic size of $Pr^{3+}$ is smaller than that of $La^{3+}$, leading to a negative differential volume ($V_{La438}$=411 Å$^3$, $V_{Pr438}$=394 Å$^3$ at room temperature). Pardo and Pickett have predicted that a smaller unit cell volume favors the metallic LS state in La438.[15] High pressure experiments by Cheng *et al*. confirmed the suppression of the semiconductor-insulator transition at $\sim$ 6 GPa ($V_{La438}$=395 Å$^3$ at 6.43 GPa),[27] although a metallic phase does not emerge, possibly due to the existence of short-range charge fluctuations. We are unable to stabilize a LS-CS state in Pr438 via DFT calculations, a result that corroborates the stability of the metallic LS phase with states around the Fermi level being largely $d_{x^2-y^2}$ in character (Supplementary Fig. 7). An alternative underpinnig for the differing behavior of Pr438 vis-à-vis La438 could be a Pr valence transition as is found in other complex oxides, with a dramatic impact on electronic and magnetic properties.[28] This possibility was tested by measuring the $M$ edge of Pr (Supplementary Fig. 8), which shows convincingly that Pr remains trivalent in the temperature range 85-300 K.

We now discuss where R438 lies with respect to the known phases of the hole-doped cuprates and nickelates. In Fig. 4 we show a schematic phase diagram of nickelates and cuprates ordered by the nominal filling of the 3d levels in the range of $d^9 \sim d^{6.9}$, with the caveat that here we are not distinguishing whether the holes are on copper or oxygen. On the bottom lie the cuprates,[1,23,29] with a superconducting dome around $d^{8.8}$. For nickelates (top part), variable hole doping in this range has been achieved for single-layer R-P phases $R_{2-x}Sr_xNiO_4$ ($R$=La, Nd),[25] with charge/spin stripe order ($d^8$~$d^{7.5}$), charge checkerboard order ($d^{7.5}$~$d^7$) and an insulator-metal transition at ~$d^7$; however, little is known between $d^9$ and $d^8$ aside from the specific cases of $La_3Ni_2O_6$ ($d^{8.5}$)[30] and $La_3Ni_2O_{6.35}$ ($d^{8.15}$),[31] which are reported to be insulators, and $LaNiO_2$ ($d^9$),[32] for which it is still unclear if the ground state is a metal or an insulator. If we superimpose the R438 system (see top of Fig. 4) on the electron concentration range appropriate to cuprates, it is located well into the overdoped, Fermi liquid side, yet is an ordered line compound with ⅓ self hole doping. Electron doping this new 'parent phase' into the concentration region where superconductivity is observed in cuprates is an obvious strategy. In addition, hole doping to move the electron count towards $d^8$ offers the opportunity to reveal potential emergent phases in the *terra incognita* at electron counts approaching $d^8$ from above. The potential role of dimensionality in such emergent phenomena can be explored in the overlap region accessed by hole doping La438 and electron doping $La_3Ni_2O_6$. Finally, the R-P trilayer La4310 exhibits a metallic ground state, sharply contrasting with the insulating behavior observed in single-layer R-P phase $La_{4/3}Sr_{2/3}NiO_4$, despite the same $3d$ electron count.[25,33] A similar situation is found for metallic bilayer $La_3Ni_2O_7$ vs insulating $La_{1.5}Sr_{0.5}NiO_4$.[25,33] Seo *et al.* have predicted charge density wave formation in $La_3Ni_2O_7$ and La4310 based on the extended Hückel tight binding method;[34] whether they exhibit any form of charge/spin order remains an open question. Recent work has also shown that $La_3Ni_2O_7$ and R4310 phases can be prepared in thin film form,[35] which offers a starting point to explore these phases, their electron/hole doped congeners, as well as pure and doped R438 phases under the influence of strain, interfacial charge transfer, and confinement.

In summary, we have shown that bulk R438 crystals possess a combination of traits that are widely considered as important ingredients for high-$T_c$ superconductivity in cuprates: a spin-½ state in which a large orbital polarization of the unoccupied $e_g$ states yields predominantly $d_{x^2-y^2}$ states at the Fermi energy, which are significantly hybridized with O 2$p$ states. Looking ahead, we note that current understanding of the cuprate phase diagram has relied on chemically doping holes or electrons into a $d^9$ parent compound, introducing quenched disorder along with carriers. Lying formally in the heavily overdoped regime of the cuprate phase diagram, the R438 system, and in particular metallic Pr438, offers a ⅓-doped parent phase that is *a priori* unperturbed by quenched disorder. These unique layered compounds thus provide a novel, complementary starting point for interrogating this correlated electron phase space and a promising set of candidates for unconventional superconductivity.

**Methods**

Methods and any associated references are available in the online version of the paper.

**References**

1  Keimer, B., Kivelson, S. A., Norman, M. R., Uchida, S. & Zaanen, J. From quantum matter to high-temperature superconductivity in copper oxides. *Nature* **518**, 179-186 (2015).

2  Tranquada, J. M. Spins, stripes, and superconductivity in hole-doped cuprates. *AIP Conf. Proc.* **1550**, 114-187 (2013).

3  Mitchell, J. F. $Sr_2IrO_4$: gateway to cuprate superconductivity? *APL Mater.* **3**, 062404 (2015).

4  Chaloupka, J. & Khaliullin, G. Orbital order and possible superconductivity in $LaNiO_3/LaMO_3$ superlattices. *Phys. Rev. Lett.* **100**, 016404 (2008).

5  Kim, Y. K. *et al.* Fermi arcs in a doped pseudospin-$^1/_2$ Heisenberg antiferromagnet. *Science* **345**, 187-190 (2014).

6  Wang, F. & Senthil, T. Twisted Hubbard model for $Sr_2IrO_4$: magnetism and possible high temperature superconductivity. *Phys. Rev. Lett.* **106**, 136402 (2011).

7  Kim, Y. K., Sung, N. H., Denlinger, J. D. & Kim, B. J. Observation of a d-wave gap in electron-doped $Sr_2IrO_4$. *Nat. Phys.* **12**, 37-41 (2016).

8   Yan, Y. J. et al. Electron-doped $Sr_2IrO_4$: an analogue of hole-doped cuprate superconductors demonstrated by scanning tunneling microscopy. *Phys. Rev. X* **5**, 041018 (2015).

9   Uchida, M. et al. Pseudogap of metallic layered nickelate $R_{2-x}Sr_xNiO_4$ (R=Nd, Eu) crystals measured using angle-resolved photoemission spectroscopy. *Phys. Rev. Lett.* **106**, 027001 (2011).

10  Anisimov, V. I., Bukhvalov, D. & Rice, T. M. Electronic structure of possible nickelate analogs to the cuprates. *Phys. Rev. B* **59**, 7901-7906 (1999).

11  Lee, K. W. & Pickett, W. E. Infinite-layer $LaNiO_2$: $Ni^+$ is not $Cu^{2+}$. *Phys. Rev. B* **70**, 165109 (2004).

12  Zhang, J. et al. Stacked charge stripes in the quasi-2D trilayer nickelate $La_4Ni_3O_8$. *Proc. Natl. Acad. Sci. U.S.A.* **113**, 8945-8950 (2016).

13  Lacorre, P. Passage from T-type to T'-type arrangement by reducing $R_4Ni_3O_{10}$ to $R_4Ni_3O_8$ (R = La, Pr, Nd). *J. Solid State Chem.* **97**, 495-500 (1992).

14  Poltavets, V. V. et al. Bulk magnetic order in a two-dimensional $Ni^{1+}/Ni^{2+}$ ($d^9/d^8$) nickelate, isoelectronic with superconducting cuprates. *Phys. Rev. Lett.* **104**, 206403 (2010).

15  Pardo, V. & Pickett, W. E. Pressure-induced metal-insulator and spin-state transition in low-valence layered nickelates. *Phys. Rev. B* **85**, 045111 (2012).

16  Pardo, V. & Pickett, W. E. Quantum confinement induced molecular correlated insulating state in $La_4Ni_3O_8$ *Phys. Rev. Lett.* **105**, 266402 (2010).

17  Botana, A. S., Pardo, V., Pickett, W. E. & Norman, M. R. Charge ordering in $Ni^{1+}/Ni^{2+}$ nickelates: $La_4Ni_3O_8$ and $La_3Ni_2O_6$. *Phys. Rev. B* **94**, 081105(R) (2016).

18  Benckiser, E. et al. Orbital reflectometry of oxide heterostructures. *Nature Mater.* **10**, 189-193 (2011).

19  Hawthorn, D. G. et al. Resonant elastic soft x-ray scattering in oxygen-ordered $YBa_2Cu_3O_{6+\delta}$. *Phys. Rev. B* **84**, 075125 (2011).

20  Kuiper, P. et al. Polarization-dependent nickel 2p x-ray-absorption spectra of $La_2NiO_{4+\delta}$. *Phys. Rev. B* **57**, 1552-1557 (1998).

21  Disa, A. S., Walker, F. J., Ismail-Beigi, S. & Ahn, C. H. Research update: orbital polarization in $LaNiO_3$-based heterostructures. *APL Mater.* **3**, 062303 (2015).

22  Hu, Z. *et al.* Hole distribution between the Ni 3*d* and O 2*p* orbitals in $Nd_{2-x}Sr_xNiO_{4-\delta}$. *Phys. Rev. B* **61**, 3739-3744 (2000).

23  Cooper, R. A. *et al.* Anomalous criticality in the electrical resistivity of $La_{2-x}Sr_xCuO_4$. *Science* **323**, 603-607 (2009).

24  The resistivity of $Pr_4Ni_3O_8$ is 67.2 Ohm cm at 298 K, and decreases with decreasing of temperature. The high resistivity could result from poor connectivity across a strain-induced network of microcracks created during the reduction process, similar to the situation reported for of $La_4Ni_3O_8$ (Ref. 12). Despite this high resistivity, its temperature dependence is that of a metal.

25  Kajimoto, R., Ishizaka, K., Yoshizawa, H. & Tokura, Y. Spontaneous rearrangement of the checkerboard charge order to stripe order in $La_{1.5}Sr_{0.5}NiO_4$. *Phys. Rev. B* **67**, 014511 (2003).

26  Uchida, M. *et al.* Pseudogap-related charge dynamics in the layered nickelate $R_{2-x}Sr_xNiO_4$ (*x*~1). *Phys. Rev. B* **86**, 165126 (2012).

27  Cheng, J. G. *et al.* Pressure effect on the structural transition and suppression of the high-spin state in the triple-layer T'-$La_4Ni_3O_8$. *Phys. Rev. Lett.* **108**, 236403 (2012).

28  García-Muñoz, J. L. *et al.* Valence transition in (Pr,Ca)$CoO_3$ cobaltites: charge migration at the metal-insulator transition. *Phys. Rev. B* **84**, 045104 (2011).

29  Gozar, A. *et al.* High-temperature interface superconductivity between metallic and insulating copper oxides. *Nature* **455**, 782-785 (2008).

30  Poltavets, V. V., Greenblatt, M., Fecher, G. H. & Felser, C. Electronic properties, band structure, and Fermi surface instabilities of $Ni^{1+}/Ni^{2+}$ nickelate $La_3Ni_2O_6$, isoelectronic with superconducting cuprates. *Phys. Rev. Lett.* **102** (2009).

31  Poltavets, V. V., Lokshin, K. A., Egami, T. & Greenblatt, M. The oxygen deficient Ruddlesden–Popper $La_3Ni_2O_{7-\delta}$ ($\delta$=0.65) phase: structure and properties. *Mater. Res. Bull.* **41**, 955-960 (2006).

32  Hayward, M. A., Green, M. A., Rosseinsky, M. J. & Sloan, J. Sodium hydride as a powerful reducing agent for topotactic oxide deintercalation: synthesis and characterization of the nickel(I) oxide $LaNiO_2$. *J. Am. Chem. Soc.* **121**, 8843-8854 (1999).

33  Ling, C. D., Argyriou, D. N., Wu, G. & Neumeier, J. J. Neutron diffraction study of $La_3Ni_2O_7$: structural relationships among $n$=1, 2, and 3 phases $La_{n+1}Ni_nO_{3n+1}$. *J. Solid State Chem.* **152**, 517-525 (2000).

34  Seo, D. K., Liang, W., Whangbo, M. H., Zhang, Z. & Greenblatt, M. Electronic band structure and madelung potential study of the nickelates $La_2NiO_4$, $La_3Ni_2O_7$, and $La_4Ni_3O_{10}$. *Inorg. Chem.* **35**, 6396-6400 (1996).

35  Lee, J. H. *et al.* Dynamic layer rearrangement during growth of layered oxide films by molecular beam epitaxy. *Nat. Mater.* **13**, 879-883 (2014).

**Acknowledgements**

Crystal growth, characterization, and theoretical calculations were supported by the U.S. Department of Energy, Office of Science, Basic Energy Sciences, Materials Science and Engineering Division. V.P. acknowledges support from Xunta de Galicia via EM2013/037 and MINECO through MAT2013-44673-R and Ramon y Cajal Program under Grant no. RyC2011-09024. ChemMatCARS Sector 15 is supported by the National Science Foundation under grant number NSF/CHE-1346572. Use of the Advanced Photon Source at Argonne National Laboratory was supported by the U.S. Department of Energy, Office of Science, Office of Basic Energy Sciences, under Contract No. DE-AC02-06CH11357. The authors thank Dr. Yu-Sheng Chen for his help with single crystal x-ray diffraction at 15-ID-B, Dr. S. Lapidus for his help with the high-resolution x-ray powder diffraction at 11-BM, and Drs. W. E. Pickett, Yu-Sheng Chen, and Y. Ren for helpful discussions.

**Author contributions**

J.F.M. and J.Z. directed the project. J.Z. and H.Z. grew single crystals. J.Z. and D.P. performed the transport measurements. J.Z. performed the powder and single crystal synchrotron x-ray diffraction experiments. J.W.F. and J.Z. performed the XAS experiments. J.W.F., J.Z. and M.R.N analyzed data. A.S.B. and V.P. performed DFT calculations. J.Z., A.S.B., J.W.F and J.F.M wrote the manuscript, with contributions from all coauthors.

**Additional information**

Supplementary information is available in the online version of the paper. Reprints and permissions information is available online at www.nature.com/reprints. Correspondence and requests for materials should be addressed to [junjie@anl.gov](junjie@anl.gov) or [junjie.zhang.sdu@gmail.com](junjie.zhang.sdu@gmail.com).

**Competing financial interests**

The authors declare no competing financial interests.

**METHODS**

**High pressure single crystal growth**. Single crystal growth of $R_4Ni_3O_{10}$ (R=La, Pr) was performed in an optical-image floating zone furnace (HKZ-1, SciDre GmbH) with 20 bar $O_2$ for $La_4Ni_3O_{10}$ and 140 bar for $Pr_4Ni_3O_{10}$. A flow rate of 0.1 L/min of oxygen was maintained during growth. Feed and seed rods were counter-rotated at 30 rpm and 27 rpm, respectively, to improve zone homogeneity. The travelling speed of the seed was 4 mm/h. After ~ 30 hours of growth, the zone and boule were quenched by shutting down the lamp. $La_4Ni_3O_8$ and $Pr_4Ni_3O_8$ single crystals (1~2 mm$^2$ × 0.5 mm) were obtained by reducing specimens cleaved from a boule of $La_4Ni_3O_{10}$ and $Pr_4Ni_3O_{10}$, respectively (flowing 4% $H_2$/Ar gas, 350 °C, five days).

**Powder x-ray diffraction**. To confirm the phase purity, powder X-ray diffraction data of $R_4Ni_3O_{10}$ and $R_4Ni_3O_8$ (R=La, Pr) were collected on a PANalytical X'Pert powder diffractometer with Cu $K_\alpha$ radiation ($\lambda$ =1.5418 Å) in the $2\theta$ range 20-60º at room temperature.

**Single crystal x-ray diffraction**. Single crystal X-ray diffraction data of $Pr_4Ni_3O_8$ were collected using a Bruker SMART APEX2 CCD area detector on a D8 goniometer operated with graphite-monochromated Mo $K_\alpha$ radiation ($\lambda$ = 0.71073 Å). A single crystal with dimensions of ~0.061 × 0.041 × 0.039 mm$^3$ was used to determine the structure at room temperature. To check for superlattice peaks at low temperature in $Pr_4Ni_3O_8$, single crystal x-ray diffraction data were collected with an APEX2 area detector using synchrotron radiation ($\lambda$=0.41328 Å) at Beamline 15-ID-B at the Advanced Photon Source, Argonne National Laboratory. Single crystals with dimensions approximately 10 μm on edge and exposure time of 1.0 s/0.2° were used. Data were collected at 15 K by flowing helium gas. $\Phi$-scans were used.

**Variable-temperature high-resolution powder x-ray diffraction**. High-resolution powder x-ray diffraction data were collected on pulverized $Pr_4Ni_3O_8$ crystals at beamline 11-BM (APS) in the range of $1° \leq 2\theta \leq 28°$, with a step size of 0.001° and counting time of 0.1 s.

**Transport measurements.** Electrical resistivity of $Pr_4Ni_3O_8$ single crystals was measured using the conventional four-probe method using silver paint contacts on a Quantum Design PPMS in the temperature range of 2-300 K. The current was applied in the *ab* plane.

**Heat capacity.** Heat capacity was measured in a Quantum Design Physical Properties Measurement System (PPMS) using the relaxation method under magnetic field of 0 T.

**X-ray absorption spectroscopy (XAS).** To study the electronic structure of $R_4Ni_3O_8$ and $R_4Ni_3O_{10}$ (R=La, Pr), XAS was performed at beamline 4-ID-C at the Advanced Photon Source (APS), Argonne National Laboratory. We carried out polarization dependent soft x-ray absorption spectroscopy at the *L* edge of Ni using the bulk sensitive total fluorescence yield (TFY) mode with the beam at grazing (15°) incidence. By controlling the polarization of the x-rays at the undulator source, we measured the orbital character projected in the *ab*-plane and along the *c*-axis without changing the beampath within the sample.

**Density-functional theory (DFT) calculations.** Electronic structure calculations were performed within DFT using the all-electron, full potential code WIEN2k[36] based on an augmented plane wave plus local orbital (APW+lo) basis set.[37] For $La_4Ni_3O_8$ the LDA+$U$ scheme that incorporates the on-site Coulomb repulsion $U$ and Hund's rule coupling $J_H$ for the Ni 3$d$ states has been applied since it is needed to converge the HS state. The values for $U$ and $J_H$ are 4.75 and 0.68 eV, respectively, as used in earlier work. For the structural relaxations in the low-spin charge stripe (LS-CS) state of $La_4Ni_3O_8$, we have used the Perdew-Burke-Ernzerhof version of the generalized gradient approximation (GGA).[38] For $Pr_4Ni_3O_8$, a full study on the energetics of the LS and HS states' dependence on $U$ has been performed within LDA+$U$ using different schemes.[39-41] $U$ and $J_H$ are applied to both the Ni 3$d$ and Pr 4$f$ states. Unlike $La_4Ni_3O_8$, a LS-CS state could not be obtained in $Pr_4Ni_3O_8$.

# References

36   Schwarz, K. & Blaha, P. Solid state calculations using WIEN2k. *Comput. Mater. Sci.* **28**, 259-273 (2003).

37   Sjöstedt, E., Nordström, L. & Singh, D. J. An alternative way of linearizing the augmented plane-wave method. *Solid State Commun.* **114**, 15-20 (2000).

38   Perdew, J. P., Burke, K. & Ernzerhof, M. Generalized gradient approximation made simple. *Phys. Rev. Lett.* **77**, 3865-3868 (1996).

39   Liechtenstein, A. I., Anisimov, V. I. & Zaanen, J. Density-functional theory and strong interactions: orbital ordering in Mott-Hubbard insulators. *Phys. Rev. B* **52**, R5467-R5470 (1995).

40   Petukhov, A. G., Mazin, I. I., Chioncel, L. & Lichtenstein, A. I. Correlated metals and the LDA+*U* method. *Phys. Rev. B* **67**, 153106 (2003).

41   Czyżyk, M. T. & Sawatzky, G. A. Local-density functional and on-site correlations: the electronic structure of $La_2CuO_4$ and $LaCuO_3$. *Phys. Rev. B* **49**, 14211-14228 (1994).